\documentclass[amsmath,amssymb]{nature}

\usepackage{color}
\usepackage{amsmath}
\usepackage{bm}
\newcommand{\new}[1]{\textcolor{black}{#1}}
\newcommand{\rev}[1]{\textcolor{black}{#1}}

\usepackage{graphicx}
\makeatletter
\let\saved@includegraphics\includegraphics
\AtBeginDocument{\let\includegraphics\saved@includegraphics}
\renewenvironment*{figure}{\@float{figure}}{\end@float}
\makeatother

\title{Large magneto-optical Kerr effect and imaging of magnetic \new{octupole} domains in an antiferromagnetic metal}

\author{Tomoya Higo $^{1,2}$, Huiyuan Man $^{1}$, Daniel B. Gopman $^{3}$, Liang Wu $^{4-6}$, Takashi Koretsune $^{2,7,8}$, Olaf M. J. van 't Erve $^{9}$, Yury P. Kabanov $^{3,10}$, Dylan Rees $^{4,5}$, Yufan Li $^{11}$, Michi-To Suzuki $^{2,8}$, Shreyas Patankar $^{4,5}$, Muhammad Ikhlas $^{1,2}$,   C. L. Chien $^{11}$, Ryotaro Arita $^{2,8}$, Robert D. Shull $^{3}$, Joseph Orenstein $^{4,5}$ and Satoru Nakatsuji $^{1,2,*}$}

\begin{document}
	
\maketitle
\begin{affiliations}
\item Institute for Solid State Physics, University of Tokyo, Kashiwa, Chiba 277-8581, Japan
\item CREST, Japan Science and Technology Agency, Kawaguchi, Saitama 332-0012, Japan
\item Materials Science and Engineering Division, National Institute of Standards and Technology, Gaithersburg, MD 20899, USA
\item Department of Physics, University of California, Berkeley, California 94720, USA
\item Materials Science Division, Lawrence Berkeley National Laboratory, Berkeley, California 94720, USA
\item Department of Physics and Astronomy, University of Pennsylvania, Philadelphia, Pennsylvania 19104, USA
\item Department of Physics, Tohoku University, Sendai, Miyagi 980-8578, Japan
\item RIKEN-CEMS, Wako, Saitama 351-0198, Japan	
\item Materials Science and Technology Division, U.S. Naval Research Laboratory, Washington, DC 20375, USA
\item Institute of Solid State Physics, Russian Academy of Sciences, Chernogolovka, Moscow Region 142432, Russia
\item Department of Physics and Astronomy, Johns Hopkins University, Baltimore, MD 21218, USA
\end{affiliations}

\begin{abstract}
When a polarized light beam is incident upon the surface of a magnetic material, the reflected light undergoes a polarization rotation \cite{Kerr1877}. This magneto-optical Kerr effect (MOKE) has been intensively studied in a variety of ferro- and ferrimagnetic materials because it provides a powerful probe for electronic and magnetic properties \cite{Zvezdin,McCord} as well as for various applications including magneto-optical recording \cite{Mansuripur}. Recently, there has been a surge of interest in antiferromagnets (AFMs) as prospective spintronic materials for high-density and ultrafast memory devices, owing to their vanishingly small stray field and orders of magnitude faster spin dynamics compared to their ferromagnetic counterparts \cite{MacDonald2011,Gomonay2014,Jungwirth2016,Baltz2016,Nemec2017}. In fact, the MOKE has proven useful for the study and application of the antiferromagnetic (AF) state. Although limited to insulators, certain types of AFMs are known to exhibit a large MOKE, as they are weak ferromagnets due to canting of the otherwise collinear spin structure \cite{Kahn1969,Smol1975,Zenkov1989,Zubov1988,Eremenko}.
Here we report the first observation of a large MOKE signal in an AF metal at room temperature.
In particular, we find that despite a vanishingly small magnetization of $M \sim$0.002 $\mu_{\rm B}$/Mn, the non-collinear AF metal Mn$_3$Sn \cite{Nakatsuji2015} exhibits a large zero-field MOKE with a polar Kerr rotation angle of 20 milli-degrees, comparable to ferromagnetic metals. 
Our first-principles calculations have clarified that ferroic ordering of magnetic octupoles in the non-collinear N{\'{e}}el state \cite{MichiToSuzuki_arXiv2016} may cause a large MOKE even  in its fully compensated AF state without spin magnetization.
This large MOKE further allows imaging of the magnetic octupole domains and their reversal induced by magnetic field. The observation of a large MOKE in an AF metal should open new avenues for the study of domain dynamics as well as spintronics using AFMs.
\end{abstract}

The magneto-optical (MO) Kerr and Faraday effects in ferro- and ferrimagnets arise from the combined effects of band exchange splitting and spin-orbit interactions (SOI) \cite{Argyres1955,Erskine1973a,Oppeneer2001} and are powerful probes of the local magnetization in such materials. In the case of  antiferromagnets (AFMs), observation of the MOKE has been restricted to a certain class of insulators (e.g. orthoferrites and iron borate) \cite{Kahn1969,Smol1975,Zenkov1989,Kimel2004,Kimel2006,Zubov1988,Kalashnikova2008,Nemec2017}, which have weak ferromagnetism due to canting of the otherwise collinear N{\'{e}}el order. In the fully compensated collinear AFMs, where the MOKE is usually absent, the quadratic MO effects such as the Voigt effect can be useful to determine the N{\'{e}}el vector \cite{Saidl2017}.
	
On the other hand, recent theoretical and experimental progress has revealed that systems such as certain spin liquids and non-collinear antiferromagnets may exhibit a large Hall response in zero applied magnetic field (anomalous Hall effect or AHE) despite a vanishing magnetization \cite{Nagaosa2010,Shindou2001,Machida2010,Chen2014,Nakatsuji2015,Kiyohara2016,Nayak2016,Ikhlas2017,Nakatsuji2017}.  Because the AHE has the same symmetry requirements as the MOKE \cite{Feng2015}, it is possible that the same class of antiferromagnets may exhibit a Kerr rotation.  Thus, the recent experimental discovery of a large AHE in the non-collinear antiferromagnet Mn$_3$Sn and its soft response to a magnetic field \cite{Nakatsuji2015} give promise for a potentially large MOKE character.

Mn$_3$Sn is a hexagonal antiferromagnet (space group P6$_3$/$mmc$) \cite{Kren1975}, \new{ which has the ABAB stacking sequence of the (0001)-plane consisting of a kagome lattice of Mn magnetic moments.} Below the  N{\'{e}}el temperature $T_{\rm N}$ $\sim$430 K, the combination of inter-site AF and Dzyaloshinskii-Moriya (DM) interactions leads to an inverse triangular spin structure, namely, a 120 degree spin structure with a uniform negative vector chirality of the in-plane Mn moments because of geometrical frustration (Figure 1a) \cite{Kren1975,TomiyoshiYamaguchi1982,Nakatsuji2015}. Significantly, this three-sublattice AF state on the kagome bilayers can be viewed as ferroic ordering of cluster magnetic octupoles (Fig. 1a) \cite{MichiToSuzuki_arXiv2016}. In addition to this dominant order parameter, the moments cant slightly in the plane to produce a small net ferromagnetic (FM) moment of $\sim$0.002 $\mu_{\rm B}$/Mn along the local easy axis in the (0001)-plane, e.g. $[2\bar{1}\bar{1}0]$ and $[01\bar{1}0]$ \cite{Nakatsuji2015}. This small  moment, which corresponds to less than $10^{-3}$ of the local Mn moment, is theoretically understood to result from the competition between the DM interaction and single-ion anisotropy \cite{Nagamiya1982}. Although, as we discuss below, the sub-dominant FM order is not responsible for the AHE and MOKE, it is essential (together with concomitant weak in-plane anisotropy) for magnetic field control of the AF spin structure. This control is demonstrated, for example, by the ability to reverse the sign of the AHE  by application of a small applied field of $B \sim$15 mT within the (0001)-plane, which is sufficient to reverse the direction of the Mn moments \cite{Nakatsuji2015}.

Before presenting the main MOKE results, we shall first discuss the AHE and magnetization curve. In this study, we used as-grown single crystals with the composition of Mn$_{3.06}$Sn$_{0.94}$ (Methods) and confirmed no magnetic transition above 50 K \cite{Nakatsuji2015}. Figures 1c and 1d provide the field dependence of the Hall resistivity $\rho_{\rm H}(B)$ and magnetization $M(B)$ for $B$ $||$ $[2\bar{1}\bar{1}0]$ at room temperature (RT), respectively (Methods). \new{The observed $\rho_{\rm H}(B = 0)$ = 3.1 $\mu$$\Omega$cm is large and is equivalent to the size of the ordinary Hall effect under an external field of $\sim$100 T estimated using the Hall coefficient of $R_0$ $\sim$0.03 $\mu\Omega$cm/T \cite{Nakatsuji2015}. This sizable $\rho_{\rm H}(B = 0)$ with vanishingly small spontaneous $M$ strongly suggests another form of order rather than FM (in this case magnetic octupole) is responsible for the AHE \cite{Nakatsuji2015,MichiToSuzuki_arXiv2016,Nagaosa2010}.}

Now, we turn to our results of the MOKE measurements. Similar to the AHE experiments, we carried out field-swept measurements of the polar MOKE using the Mn$_{3}$Sn ($2\bar{1}\bar{1}0$)-plane with a $\lambda = $ 660 nm semiconductor laser at 300 K (Methods, Fig. 1b).
Significantly, a large change of the Kerr rotation angle $|\theta_{\rm K}| = 17.5$ milli-degrees and a clear square hysteresis loop are observed as a function of the field $B$ $||$ $[2\bar{1}\bar{1}0]$ (Fig. 2a).
The coercive field of $B_{\rm C}$ $\sim$12 mT is consistent with the hysteresis curve obtained in $\rho_{\rm H}(B)$, indicating that the magnetic properties at the surface are nominally identical to those in the bulk.
The MOKE in an AF metal is unprecedented.
Figure 2b provides the longitudinal MOKE loops measured in various $B$ directions within the ($2\bar{1}\bar{1}0$)-plane at 300 K (Methods). \new{Upon changing the direction of both the applied field $B$ and plane of incidence from $[01\bar{1}0]$ ($\varphi= 0^{\circ}$) to $[0001]$ ($\varphi = 90 ^{\circ}$) in the ($2\bar{1}\bar{1}0$)-plane (Fig. 2c), we find that the MOKE signal decreases and the coercivity increases. Finally, the MOKE signal disappears when $B$ and the plane of incidence are parallel to the magnetic hard axis [0001] ($\varphi = 90^{\circ}$).} The signal magnitude is proportional to the projection of $B$ onto $[01\bar{1}0]$, which demonstrates that the surface magnetic character probed by the MOKE is consistent with our previous studies on bulk transport and magnetization \cite{Nakatsuji2015}.
We also performed polar MOKE spectroscopy on the same sample with the wavelength from 400 nm to 1100 nm at 300 K (Methods). 
In Fig. 2d, we plot the spontaneous Kerr rotation $\theta_{\rm K}$($B$ = 0) = $(\theta_{\rm K, +0 T} - \theta_{\rm_K, -0 T})$/2, which removes time-reversal even effects such as birefringence resulting from structural anisotropy and the Voigt effect, and illustrates the history dependent remnant response. The spectra show a broad peak with magnitude 19.6 milli-degrees at $\lambda$ = 580 nm (2.1 eV).

A striking property of the MOKE data is that, once application of a small magnetic field serves to rotate the spin orientation, further increase of the field and therefore the magnetization, does not affect much on the size of the Kerr effect (Fig. 1d and Fig. 2a).
This observation strongly suggests that the amplitude of the MOKE and the FM magnetization are entirely decoupled in this system. As a result, the ratio of the MOKE amplitude to $M$ is unusually large. To emphasize this point, we compare the polar MOKE in Mn$_3$Sn with values reported for ferro- and ferrimagnets and AF insulators. Figure 3 is a double-logarithmic plot of the polar MOKE signal vs. magnetization (Methods)
obtained at RT. The wavelengths used in the various measurements are all in the range between \rev{250 nm} and 1250 nm. \new{For FM systems,} there is a clear trend that the Kerr angle increases with the magnetization. This is consistent with the theoretical expectation that the MOKE is roughly proportional to $M$ (and SOI) \cite{Argyres1955}. However, this correlation is not as straightforward as is the case for the AHE \cite{Nagaosa2010,Manyala2004} because the MOKE  can exhibit a complicated frequency dependence including sign changes. Nonetheless, we note that the ratio $|\theta_{\rm K}|$/$M$ = 25.6 degrees/T of Mn$_3$Sn is largest among all the empirical reports, i.e. 1-2 order(s) of magnitude larger than ferro- and ferrimagnets, and still the largest among those known for canted AF insulators that have a similar magnitude to Mn$_3$Sn.  

The experimental evidence presented above for the decoupling of the MOKE from the FM order is strongly supported by theoretical considerations. Detailed symmetry analysis \cite{MichiToSuzuki_arXiv2016} demonstrated that, despite vanishing $M$, ferroic octupole order was sufficient to induce a non-zero net Berry curvature of the Fermi sea, and hence an AHE. Further it was demonstrated that the size of the AHE in Mn$_{3}$Sn could be reproduced in a model in which the spin orientations are set such that $M=0$.  As the AHE corresponds to the $\omega \rightarrow 0$ limit of the MOKE, the same symmetry considerations apply to both. To further test the connection of this theory to experiment, we have performed a  first-principles calculation of the frequency dependence of the Kerr rotation $\theta_{\rm K}(\omega)$ for the fully compensated AF state without spin canting for Mn$_3$Sn (Methods). The calculation reproduces the large Kerr rotation of $\sim$30 milli-degrees at around 520 nm (2.4 eV) (Fig. 2d inset), which corresponds to the observed value of $\sim$20 milli-degrees at 580 nm (2.1 eV). Our results of a large MOKE $without$ spin magnetization sharply contrast with the previously known AF insulator cases where weak ferromagnetism has been believed to be essential for the presence of the MOKE \cite{Smol1975,Zenkov1989,Solovyev1997}.
In addition, we find almost no change in the Kerr angle by increasing $M$ from 0 to 0.005 ${\mu_{\rm B}}$/formula unit (f.u.) (Fig. 2d inset). In contrast, the change in the size of the octupole, which is proportional to the sublattice moment, should lead to a significant change in the band structure and AHE \cite{MichiToSuzuki_arXiv2016}, and hence in the MOKE. Therefore, in Mn$_3$Sn, it is not magnetization but magnetic octupole that induces the large MOKE.
Further reflectivity measurements using e.g. Fourier-transform infrared spectroscopy will be useful to clarify $\sigma_{ij}(\omega)$ and will help understand the origin of the MOKE in the AF metallic state.

Having shown the sizable Kerr rotation in Mn$_3$Sn, we demonstrate the application of the MOKE microscopy as a useful non-contact and non-destructive probe for imaging of magnetic domains and their corresponding dynamics. In particular, we apply the polar MOKE to visualize the magnetic domains in the AF state of Mn$_3$Sn (Methods). A series of MOKE images of the Mn$_3$Sn ($2\bar{1}\bar{1}0$)-plane under $B$ $||$ $[2\bar{1}\bar{1}0]$ (-21 mT $\leq B \leq $ 21 mT) are presented in Figs. 4a to 4h. Changes in contrast from grey-to-black and black-to-grey are evident over the sequence of increasing field (Figs. 4a-d) and decreasing field (Figs. 4e-4h), respectively. Here, the grey and black colours correspond to positive and negative values of the polar MOKE signal $||$ $[2\bar{1}\bar{1}0]$. As discussed above, the symmetry considerations show that switching of the \new{magnetic octupole} domains with a triangular spin texture can flip the sign of $\sigma_{ij}(\omega)$ and consequently the Kerr angle $\theta_{\rm K}$. Particularly in $B$ $||$ $[2\bar{1}\bar{1}0]$, one of the three moments of a Mn triangle aligns with the field, with the other two at $\pm$ 120 degrees away (Fig. 1a). Therefore, the grey and black regions are interpreted as \new{the domains in which the Mn spin collinear with the field (the magnetic octupolar axis)} points either along the positive or negative $[2\bar{1}\bar{1}0]$ direction (Fig. 4i). Provided that the in-plane Mn moments have an easy-axis along $[2\bar{1}\bar{1}0]$, Mn$_3$Sn may have six types of domains in principle, corresponding to six equivalent axes to $\langle2\bar{1}\bar{1}0\rangle$ in the (0001)-plane. However, a clear square shape observed in our MOKE hysteresis loop, AHE and magnetization measurements for $B$ $||$ $[2\bar{1}\bar{1}0]$ (Figs. 1 and 2) all suggest that the field cycle only stabilizes the two types of the magnetic octupole domains shown in Fig. 4i. Indeed, only two distinct regions with different colours are found in Figs. 4b, c, f and g.

Generally, in ferromagnets, two fundamental processes in the magnetization reversal are known, namely, (i) reversal by domain nucleation and domain wall propagation, and (ii) reversal by coherent rotation. The series of our domain images indicates the former mechanism while the coarse resolution hinders the observation of nucleation itself. At the highest field $B$ = 18.4 mT (Fig. 4d), the view of the entire region (25 $\mu$m $\times$ 50 $\mu$m) reflects an oppositely aligned domain in contrast to the initial domain configuration (Fig. 4a).
Similar evolution of the domain images is obtained for the field-increasing and -decreasing processes (Fig. 4).
Interestingly, the domain images around zero field in the field-increasing (Fig. 4a) and -decreasing processes (Fig. 4e) show grey/black contrast, respectively, confirming that these mono-domains have a spontaneous Kerr angle at zero field whose sign can be switched by a modest coercive field $\sim$10 mT. This is the first observation of the domain reversal in an AF metal \rev{and the magnetic octupole domain reversal} by MOKE microscopy.

The observation of the large MOKE signal and the \rev{magnetic octupole} domain reversal in the antiferromagnet Mn$_3$Sn has various implications. 
In spintronics, antiferromagnets have attracted significant attention as a potential active element in the next generation memory technology \cite{MacDonald2011,Gomonay2014,Jungwirth2016,Baltz2016,Nemec2017}.
The MOKE in AF metals may well accelerate the development of future memory devices with emergent optical, magnetic and electric properties by taking advantage of significantly large reflectivity in comparison with insulators.
Furthermore, our success in the magneto-optical imaging of the domain configurations in an AF metal opens a new avenue for studying current-driven domain dynamics. In principle, the AF domain dynamics could be very different from the ones in ferromagnets as the magneto-static energy, which is the primary driving force of the domain formation in ferromagnets, is unimportant in antiferromagnets. In particular in Mn$_3$Sn, six magnetic domains and the associated various types of topological defects could be possible including a vortex type of point defect around which the six domains meet. It is a future subject to investigate the possible current-induced AF domain wall motion \cite{Liu2017}.
Finally, the topological aspect of the magnetic state (magnetic Weyl state) of Mn$_3$Sn, which has been recently pointed out by theoretical \cite{Yang2016} and experimental \cite{Kuroda2017} studies, would make it interesting to investigate the Berry curvature effects in the MOKE at low frequency, and further make the topological defects even more attractive as the bulk-edge correspondence for such a magnetic Weyl semimetallic state may lead to a Fermi arc in a magnetic domain wall, and contribute to the MOKE signal in an unconventional fashion \cite{Liu2017,Trivedi2015}. 

\newpage

\begin{methods}
\subsection{Sample preparation and characterization.}
Polycrystalline samples were prepared by melting the mixtures of manganese and tin in an alumina crucible sealed in an evacuated quartz ampoule in a box furnace at 1050 $^\circ$C for 6 hours. Excess manganese (10 mol.\%) over the stoichiometric amount was added to compensate for its loss during crystal growth. The obtained polycrystalline materials were used for growing single crystals by the Bridgman method using a single zone furnace with a maximum temperature of 1080 $^\circ$C and growth speed of 1.5 mm/h. Our single-crystal and powder x-ray measurements indicate the presence of only single phase hexagonal Mn$_3$Sn with the lattice constants of $a$ = 5.66(1) \AA~and $c$ = 4.53(1) \AA. Our scanning electron microscopy energy dispersive x-ray spectroscopy (SEM-EDX) and inductively coupled plasma atomic emission spectrometry (ICP-AES) analyses confirmed that Mn$_3$Sn is the bulk phase, and found that the composition of the single crystals is Mn$_{3. 06}$Sn$_{0.94}$.
The magnetization was measured using a commercial SQUID magnetometer (MPMS, Quantum Design). The Hall resistivity was measured by a standard four-probe method using a commercial physical property measurement system (PPMS, Quantum Design). Our as-grown single crystals are found to exhibit a phase transition to a cluster glass phase at 50 K \cite{Nakatsuji2015}. In this study, we used as-grown single crystals that show no transition above 50 K.

\subsection{Magnetic field dependence of the magneto-optical Kerr effect (MOKE) shown in Figs. 2a, 2b, \& 2c.}
Magnetic field dependence of the Magneto-optical Kerr effect (MOKE) was measured by using a commercial system (NanoMOKE3, Quantum Design).
Bulk Mn$_3$Sn crystals were used for the MOKE spectroscopy measurement. Crystal samples were cut and polished so that the samples had optically smooth surfaces along the ($2\bar{1}\bar{1}0$)-plane. Kerr rotation versus applied field were measured in a polar MOKE configuration under out-of-plane applied magnetic fields, and in a longitudinal MOKE configuration under in-plane applied magnetic fields. \new{In the polar MOKE configuration, a polarized light beam and magnetic field are applied perpendicular to the reflecting ($2\bar{1}\bar{1}0$)-plane. \new{The plane of the input polarization is parallel to the [0001] direction.
In the longitudinal MOKE configuration, the magnetic field is parallel to both the plane of incidence and the reflecting surface. The polarized light beam is applied under an incident angle of 45 degrees and the plane of polarization is parallel to the plane of the incidence. $\varphi$ is the angle between the plane of incidence/magnetic field and the $[01\bar{1}0]$ direction.}} An electromagnet was used to generate both in-plane and out-of-plane applied magnetic fields up to 0.24 T. Fast MOKE loops were acquired using a 660 nm semiconductor laser and a spatial light modulator (SLM) enabling acquisition of 30 hysteresis loops per minute while the applied field was sinusoidally varied at a rate of 0.5 Hz. One hundred MOKE loops were averaged in order to improve the signal to noise fidelity. \new{To obtain the MOKE signal, which is an odd function of magnetic field $B$ (more precisely magnetization $M$ and/or ferroic octupole order parameter in Mn$_3$Sn), we calculated the Kerr rotation to be $\{\theta_{\rm K}(+B) - \theta_{\rm K}(-B)\}$/2.}

\subsection{Magneto-optical Kerr effect (MOKE) spectroscopy shown in Fig. 2d.}
Magneto-optical Kerr effect spectroscopy was performed by using a W-filament light source and grating monochromator. The ($2\bar{1}\bar{1}0$)-polished Mn$_3$Sn single crystal was used for the measurement as with the MOKE hysteresis loop measurement.
In a polar Kerr measurement, linearly polarized light was focused onto the sample with a $\sim$10 $\mu$m diameter spot by a reflective objective with an incidence angle close to normal and the reflected light was collected by a pick-up mirror. To measure the Kerr rotation, we used a technique based on a photoelastic optical phase modulator (PEM) \cite{Badoz1977,Patankar2015}. Synchronous detection of the reflected light at the second harmonic of the PEM frequency and DC reflectivity enabled precision of rotation with 0.25 milli-degrees (5 $\mu$rad) sensitivity.
\new{The plane of the input polarization is parallel to the [0001] direction. Spectra were acquired for the following sequence of fields: $+0.2$ T (with respect to $[2\bar{1}\bar{1}0]$), $+0$ T, $-0.2$ T, and then finally $-0$ T again. We refer to the zero field condition following $\pm0.2$ T as $\pm0$ T.
The spectra exhibit the same hysteretic and remnant behaviour as the AHE and single wavelength MOKE observations.
In particular, Kerr rotation is observed at 0.2 T, remains unchanged at $+0$ T, flips sign when measured at $-0.2$ T, and then remains at $-0$ T with opposite sign to the MOKE signal at $+0$ T.
In Fig. 2d, we plot the spontaneous magneto-optical Kerr rotation $\theta_{\rm K}$($B$ = 0) = $(\theta_{\rm K, +0 T} - \theta_{\rm K, -0 T})$/2, where $\theta_{\rm K, \pm0 T}$ is the Polar Kerr angle at $\pm$0 T.  This estimate of $\theta_{\rm K}$($B$ = 0) removes time-reversal even effects such as birefringence resulting from structural anisotropy and the Voigt effect}, and illustrates the history dependent remnant response.

\subsection{Magneto-optical Kerr effect (MOKE) imaging shown in Fig. 4.}
For magneto-optical imaging of the magnetic domain configurations in bulk Mn$_3$Sn crystals, the same ($2\bar{1}\bar{1}0$)-polished sample, which was used in the MOKE hysteresis loop and spectroscopy measurements, was used. The magneto-crystalline anisotropy tends to align the magnetization along the [$2\bar{1}\bar{1}0$] direction, which enables us to conduct polar MOKE investigation of the out-of-plane magnetic domains in the ($2\bar{1}\bar{1}0$)-plane. Magneto-optical imaging was performed at room temperature using a polarizing microscope with nearly-crossed polarizers and a 625 nm collimated LED source. Direct visualization of out-of-plane domain structures was provided during the domain-reversal process using the polar magneto-optical Kerr effect. The black and grey colours of the magneto-optical image correspond to opposite signs of the magnetic domain direction. Magneto-optical imaging was made using a charge coupled device (CCD) camera. Applied out-of-plane magnetic fields were generated by a projected-field electromagnet, enabling the application of magnetic fields up to 21 mT (16.8 kA/m) in magnitude.
Certain equipment, instruments or materials are identified in this paper in order to adequately specify the experimental details. Such identification does not imply recommendation by the endorsement by the authors or their organizations nor does it imply the materials are necessarily the best available for the purpose.

\subsection{Computational details.}
Under the magnetic structure shown in Fig. 1a, the conductivity tensor has the form
\begin{align}
\bm \sigma = \begin{pmatrix}
\sigma_{xx} & 0 & 0 \\
0 & \sigma_{yy} & \sigma_{yz} \\
0 & -\sigma_{yz} & \sigma_{zz}
\end{pmatrix},
\label{eq:conductivity}
\end{align}
where $x$, $y$, and $z$ correspond to [$2\bar{1}\bar{1}0$], [$01\bar{1}0$], and [0001] directions, respectively.
Here, $\sigma_{\alpha\beta}$ is approximated using the Kubo formula as
\begin{align}
\sigma_{\alpha\beta}(\hbar \omega) = \frac{i e^2 \hbar}{N_k \Omega_c} \sum_{\bm k, n,m}
\frac{f(\epsilon_{m\bm k}) - f(\epsilon_{n\bm k})}{\epsilon_{m\bm k} - \epsilon_{n\bm k}}
\frac{ \langle \psi_{n \bm k} | v_\alpha | \psi_{m \bm k} \rangle  \langle \psi_{m \bm k} | v_\beta | \psi_{n \bm k} \rangle }
{ \epsilon_{m\bm k} - \epsilon_{n\bm k} - (\hbar \omega + i \eta) },
\label{eq:cond_def}
\end{align}
where $\Omega_c$ is the cell volume, $N_k$ is the number of $k$-points, $f(\epsilon)$ is the Fermi distribution function, and $\eta$ is the smearing parameter.
Since the incident light is along the $x$-axis, the polar Kerr effect can be described by the two eigen modes, $\bm E_{\pm}$ and corresponding complex refractive indices, $n_{\pm}$. 
Here, $\bm E_{\pm}$ and $n_{\pm}$ are the solutions of the Fresnel equation for Eq.\ \eqref{eq:conductivity} and given as
\begin{align}
n_{\pm}^2 = 1 + \frac{4 \pi i}{\omega} \left( \frac{\sigma_{yy} + \sigma_{zz}}{2} \pm \sqrt{ \frac{1}{4}(\sigma_{yy} - \sigma_{zz})^2 - \sigma_{yz}^2 } \right).
\end{align}
Then, by decomposing the linearly polarized incident light, $\bm E_z$, as
$\bm E_z = p \bm E_+ + q \bm E_-$, the reflected light is written as
\begin{align}
p \frac{n_+ - 1}{n_+ + 1} \bm E_+ + q \frac{n_- - 1}{n_- + 1} \bm E_-.
\end{align}
Thus, by analyzing the orientation of the ellipse of this light, the Kerr rotation, $\theta_{\rm K}$ is obtained.
For calculating Eq.\ \eqref{eq:cond_def},
the Kohn-Sham energy, $\epsilon_{n \bm k}$, and wavefunction, $| \psi_{n \bm k} \rangle$, were obtained within the generalized-gradient approximation \cite{Perdew1996} based on the density functional theory as implemented in the quantum-{\sc ESPRESSO} package \cite{Giannozzi2009}.
A 7 $\times$ 7 $\times$ 7 $k$-point grid, ultrasoft pseudopotentials \cite{Vanderbilt1990} and plane wave basis sets with cutoff energies of 80 Ry for wavefunctions and 320 Ry for charge densities were used.
The obtained magnetic moment for each Mn atom is 3.26 $\mu_{\rm B}$ while the total magnetic moment vanishes due to cancellation within the numerical error.
We also performed calculation with constraint on $M$ and calculated $\theta_{\rm K}$ for $M$ = 0.005 and 0.025 $\mu_{\rm B}$/f.u..
The summation in Eq.\ \eqref{eq:cond_def} were taken using a Wannier-interpolated band structure \cite{Mostofi2008} with a 50 $\times$ 50 $\times$ 50 $k$-point grid and a smearing width of $\eta = 0.4$ eV.

\subsection{Detailed information about MOKE results for \new{ferro- and ferrimagnets, and antiferromagnetic insulators} shown in Fig. 3.}
Figure 3 is made using the polar MOKE results obtained at room temperature (RT) from \new{various (i) ferro- and ferrimagnetic metals, (ii) ferro- and ferrimagnetic insulators, and (iii) antiferromagnetic insulators}, as reported in previous studies including (i) L1$_0$-Mn$_{x}$Ga$_{1-x}$: $0.43 \leq x \leq 0.56$ ($\lambda$ = 670 nm, $T$ = 300 K) \cite{L10MnGa_M_Zhu2013,L10MnGa_K_Zhu2016} and $x$ = 0.62 (820 nm, RT) \cite{MnGa_Krishnan1992}, L1$_2$-Mn$_3$Pt (650 and 1000 nm, RT) \cite{L12MnPt3_Kato1995}, MnBi (700 and 900 nm, RT) \cite{MnBi_Di1994}, MnSb$_{1-x}$Sn$_{x}$: $0 \leq x \leq 0.4$ (633 nm, RT) \cite{Mn2SbSn_Yamamoto1991}, and MnAlGe (633 nm, RT) \cite{Buschow1983}, PtMnSb (710 nm, RT) \cite{PtMnSb_Ohyama1987}, Ni (633 nm, RT) \cite{Buschow1983}, Co (633 nm, RT) \cite{Buschow1983}, Co$_x$Ga$_{1-x}$: $0.58 \leq x \leq 0.7$ (633 nm, RT) \cite{Buschow1983}, Co$_2$Sn (633 nm, RT) \cite{Buschow1983}, Fe (633 nm, RT) \cite{Buschow1983}, Fe$_x$Al$_{1-x}$: $0.7 \leq x \leq 0.9$ (633 nm, RT) \cite{Buschow1983}, Fe$_x$Ga$_{1-x}$: $0.6 \leq x \leq 0.8$ (633 nm, RT) \cite{Buschow1983}, Fe$_x$Ge$_{1-x}$: $x$ = 0.85 and 0.9 (633 nm, RT) \cite{Buschow1983}, D0$_{19}$-Fe$_3$Ge (633 nm, RT) \cite{Buschow1983}, D0$_{19}$-Fe$_3$Sn (633 nm, RT) \cite{Buschow1983}, MnSb$_x$Sn$_{1-x}$: $0 \leq x \leq 0.4$ (633 nm, RT) \cite{Mn2SbSn_Yamamoto1991}, and L1$_0$-FePt (633 nm, RT) \cite{L10FePt_Mitani1995}, \new{ and (ii) Fe$_3$O$_4$ (1240 nm, RT) \cite{Zhang1981}, Y$_3$Fe$_5$O$_{12}$ (251 nm, RT) \cite{Kahn1969,Gilleo1980}, Eu$_3$Fe$_5$O$_{12}$ (253 nm, RT) \cite{Kahn1969-2,Gilleo1980}, Ho$_3$Fe$_5$O$_{12}$ (279 nm, RT) \cite{Visnovsky1981,Gilleo1980}, and Er$_3$Fe$_5$O$_{12}$ (251 nm, RT) \cite{Kahn1969,Gilleo1980}, and (iii) YFeO$_3$ (421 nm , RT) \cite{Kahn1969,Nielsen1971}, SmFeO$_3$ (430 nm, RT) \cite{Kahn1969,Treves1965}, EuFeO$_3$ (432 nm, RT) \cite{Kahn1969,Nielsen1971}, DyFeO$_3$ (324 nm, RT) \cite{Kahn1969,Nielsen1971}, ErFeO$_3$ (320 nm, RT) \cite{Kahn1969,Treves1965}, YbFeO$_3$ (328 nm, RT) \cite{Kahn1969,Treves1965}, LuFeO$_3$ (324 nm, RT) \cite{Kahn1969,Treves1965}.}
\subsection{}
The data that support the plots within this paper and other findings of this study are available from the corresponding author upon reasonable request.
\end{methods}



\section*{References}


\begin{addendum}
\item We thank Y. Otani, M. Kimata, L. Balents, H. Chen, H. Ishizuka, O. Tchernyshyov and C. Broholm for useful discussions.
This work is partially supported by CREST(JPMJCR15Q5), Japan Science and Technology Agency, by Grants-in-Aids for Scientific Research on Innovative Areas (15H05882 and 15H05883) from the Ministry of Education, Culture, Sports, Science, and Technology of Japan, and by Grants-in-Aid for Scientific Research (16H02209) and Program for Advancing Strategic International Networks to Accelerate the Circulation of Talented Researchers (No. R2604) from the Japanese Society for the Promotion of Science (JSPS). Kerr spectroscopy was performed at Lawrence Berkeley National Laboratory in the Spin Physics program supported by the Director, Office of Science, Office of Basic Energy Sciences, Materials Sciences and Engineering Division, of the U.S. Department of Energy under Contract No. DEAC02-05CH11231. L.W. is supported by the Gordon and Betty Moore Foundation's EPiQS Initiative through Grant GBMF4537 to J.O. at UC Berkeley.
Y.L. has been supported in part by SHINES, grant SC0012670, an Energy Frontier Research Center of US Department of Energy and grant DE-SC0009390. This research is funded in part by a QuantEmX grant from ICAM and the Gordon and Betty Moore Foundation through Grant GBMF5305. The work of T.H., H.M., and S.N. at IQM was partially supported by the US Department of Energy, office of Basic Energy Sciences, Division of Material Sciences and Engineering under grant DE-FG02-08ER46544. T.H., H.M. and S.N. greatly appreciate the hospitality of Department of Physics and Astronomy of Johns Hopkins University, where a part of this work was conducted. The use of the facilities of the Materials Design and Characterization Laboratory at the Institute for Solid State Physics, The University of Tokyo, is gratefully acknowledged.

\item[Author Contributions]
S.N. conceived the project. S.N., R.D.S., J.O., and C.L.C. planned the experiments. M.I. synthesized the single crystals. T.H., M.I., and S.N. performed magnetization and Hall effect measurements. T.H. and H.M. prepared the samples for magneto-optical experiments. O.M.J.E. and D.B.G. performed the MOKE loop experiment. L.W., D.R., and S.P. performed the MOKE spectroscopy experiment. T.H., H.M., D.B.G., and Y.P.K. performed the MOKE imaging experiment. Y.P.K. and Y.L. carried out the image processing. R.A. planned the theoretical calculations, and T.K., M.-T.S. and R.A. performed the first-principles calculations. T.H., D.B.G., L.W., O.M.J.E., C.L.C., R.A., R.D.S., J.O., and S.N. discussed the results and T.H., D.B.G., L.W., T. K., R.A., J.O. and S.N. wrote the manuscript and prepared figures; All authors commented on the manuscript.

\item[Competing Interests]
The authors declare that they have no competing financial interests.

\item[Correspondence] 
Correspondence and requests for materials should be addressed to S.N.\\
(email: satoru@issp.u-tokyo.ac.jp).

\end{addendum}

\newpage

\begin{figure}
\caption{{\bf Crystal and magnetic structures, anomalous Hall effect and weak ferromagnetism at 300 K of the non-collinear antiferromagnet Mn$_3$Sn.}
{\bf a,} \rm Large red (small dark grey) and large \new{transparent} orange (small transparent grey) spheres represent Mn (Sn) atoms forming kagome planes at $z$ = 0 and 1/2, respectively. The Mn magnetic moments (arrows) \new{lie in the (0001)-plane} and form an inverse triangular spin structure. \new{Different colours of arrows indicate three sublattices. The spin structure on the kagome bilayers can be considered as ferroic ordering of cluster magnetic octupoles by using the cluster octupole unit shown in the inset.} 
{\bf b,} Schematic illustration of magneto-optical Kerr effect measurements. A polarized light beam is applied perpendicular to the ($2\bar{1}\bar{1}0$)-plane, and the reflected light becomes elliptically polarized with the major axis rotated by $\theta_{\rm K}$.
{\bf c,d} Field dependence of {\bf c,} the Hall resistivity $\rho_{\rm H}$ in the magnetic field $B$ $||$ $[2\bar{1}\bar{1}0]$ with the electric current $I$ $||$ $[0001]$, and of {\bf d,} the magnetization $M$ in $B$ $||$ $[2\bar{1}\bar{1}0]$.}
\end{figure}

\begin{figure}
\caption{{\bf Magneto-optical Kerr rotation in the non-collinear antiferromagnet Mn$_3$Sn at 300 K.}
\rm Field dependence of {\bf a,} the polar magneto-optical Kerr rotation angle $\theta_{\rm K}$ for the  ($2\bar{1}\bar{1}0$)-plane in $B$ $||$ $[2\bar{1}\bar{1}0]$, and 
{\bf b,} \rm the longitudinal magneto-optical Kerr rotation angle $\theta_{\rm K}$ of the ($2\bar{1}\bar{1}0$)-plane measured along $\varphi$ = 0-90 degrees (Methods). 
{\bf c,} \rm Schematic illustrations of sample configurations for the polar and longitudinal magneto-optical Kerr effects (MOKE) (Methods). 
{\bf d,} \rm Polar MOKE spectroscopy of the ($2\bar{1}\bar{1}0$)-plane from 400 nm to 1100 nm under zero magnetic field (Methods).
\new{Inset: Polar Kerr rotation spectrum obtained using the first-principles calculation with the smearing parameter of $\eta = 0.4$ eV (Methods).} The rotation angle slightly changes with $M$, and there is almost no change in the angle between $M$ = 0 and 0.005 $\mu_{\rm B}$/f.u. Since the observed $M$ is just $\sim$0.005 $\mu_{\rm B}$/f.u. ($\sim$0.002 $\mu_{\rm B}$/Mn), we may conclude that $M$ plays just a minor role in the MOKE.}
\end{figure}
	
\begin{figure}
\caption{{\bf Magnetization dependence of the polar Kerr effect at room-temperature for \new{ferro- and ferrimagnets, and antiferromagnets including Mn$_3$Sn}.}
\rm Full logarithmic plot of the polar magneto-optical Kerr rotation angle $|\theta_{\rm K}|$ vs. the magnetization $M$ for various \new{ferro- and ferrimagnets, and antiferromagnets including Mn$_3$Sn} measured around room-temperature (Methods). \new{Here, Mn$_3$Sn is the only non-collinear antiferromagnet even in its fully compensated antiferromagnetic state.} In ferromagnetic systems, it shows a general trend for most of the ferro- and ferrimagnets that $|\theta_{\rm K}|$ increases with increasing $M$, namely $|\theta_{\rm K}| = K_{\rm S} M$. Here, $K_{\rm S}$ is a coefficient, ranging between 0.2 and 2.0 degrees/T for ferro- and ferrimagnets, as highlighted by the shaded region. The spontaneous polar MOKE signal of Mn$_3$Sn does not follow the above scaling for ferromagnets and $K_{\rm S}$ has a 1-2 order(s) of magnitude larger value of 25.6 degrees/T, \new{which is still the largest among antiferromagnets that have similar magnitude of $K_{\rm S}$ = 10-20 degrees/T}.}
\end{figure}
	
\begin{figure}
\caption{{\bf Magneto-optical Kerr effect (MOKE) images in the non-collinear antiferromagnet Mn$_3$Sn.}
\rm Evolution of the antiferromagnetic domains of the ($2\bar{1}\bar{1}0$)-plane as a function of a field $B$ along $[2\bar{1}\bar{1}0]$ (-21 mT $\leq B \leq $ 21 mT). The imaging area is 25 $\mu$m $\times$ 50 $\mu$m. Grey and black regions correspond to positive and negative values of the MOKE signal. {\bf a}-{\bf d,} MOKE images obtained  in the increasing field process from -21 mT to 21 mT. {\bf e}-{\bf h,} MOKE images obtained in the decreasing field process from 21 mT to -21 mT. ({\bf i}) Schematic illustration of two regions with different MOKE image contrasts (grey/black areas) due to opposite signs of the Kerr angles $\theta_{\rm K, -/+}$, corresponding to two types of \new{cluster magnetic octupule} domains which have the inverse triangular spin structures with opposite spin directions within the  (0001)-plane. The two regions should be separated by a domain wall (yellow area).}
\end{figure}

\begin{figure}
\begin{center}
\includegraphics[width=\columnwidth]{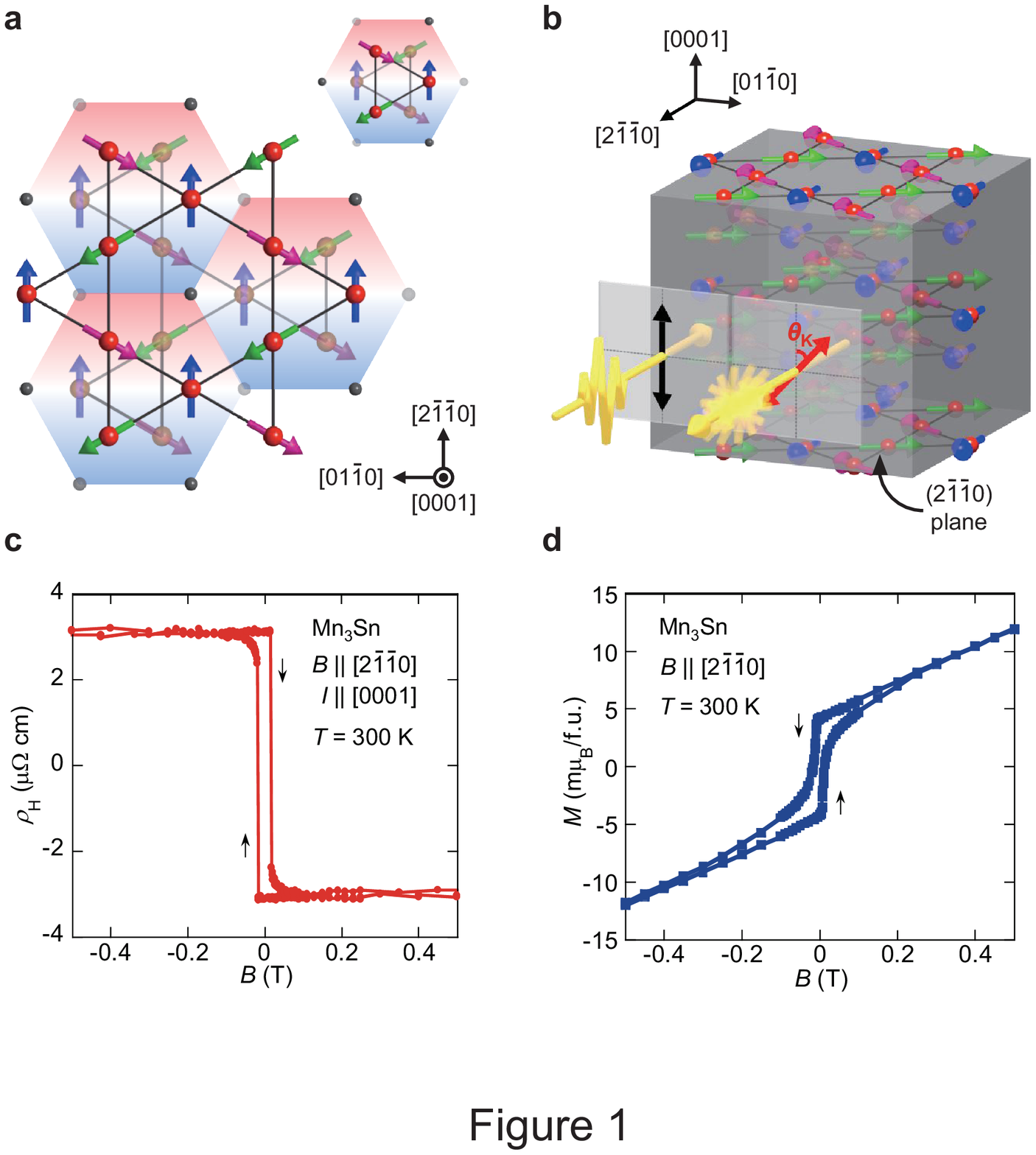}
\end{center}
\end{figure}

\begin{figure}
\begin{center}
\includegraphics[width=14cm]{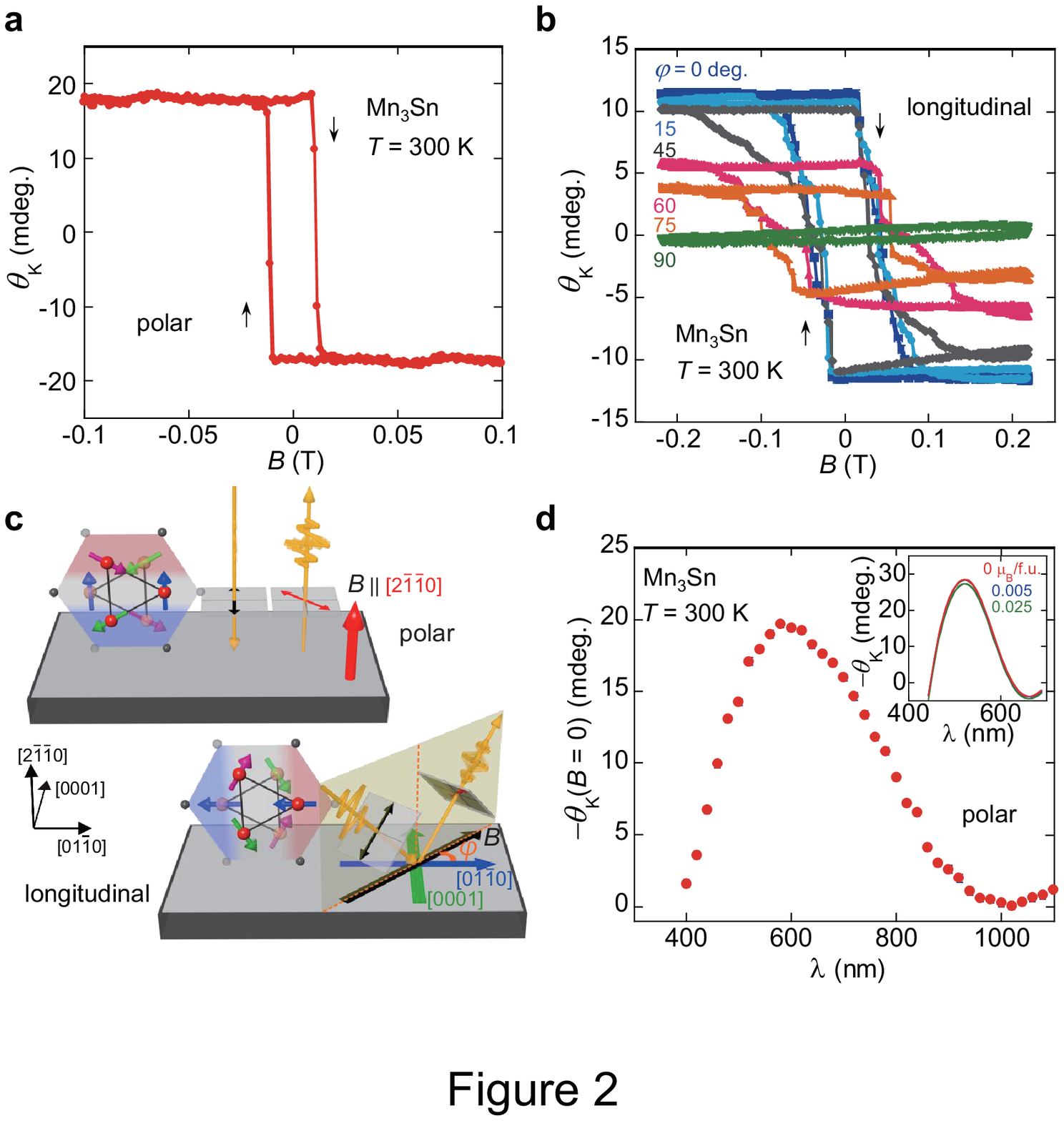}
\end{center}
\end{figure}

\begin{figure}
\begin{center}
\includegraphics[width=\columnwidth]{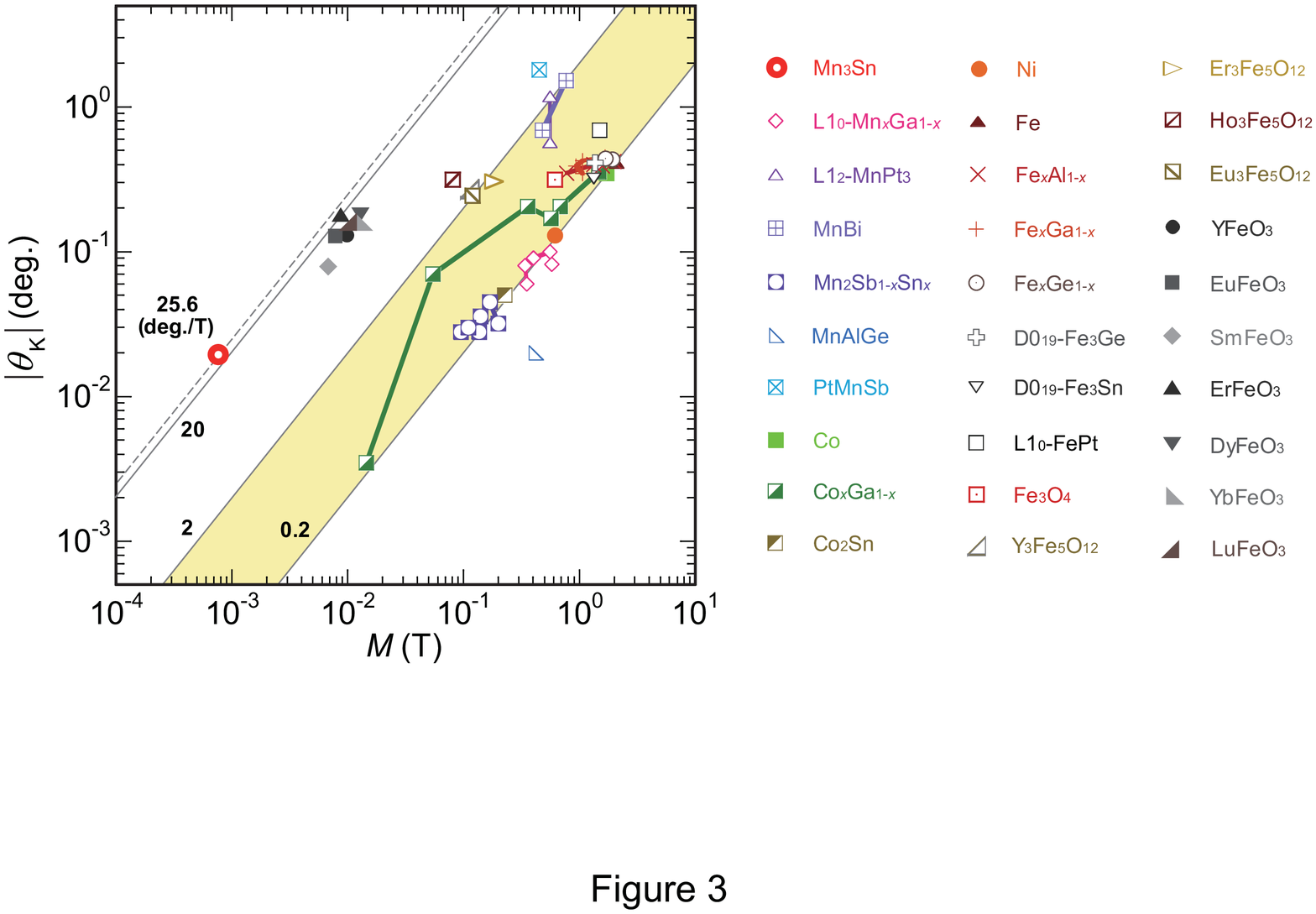}
\end{center}
\end{figure}

\begin{figure}
\begin{center}
\includegraphics[width=\columnwidth]{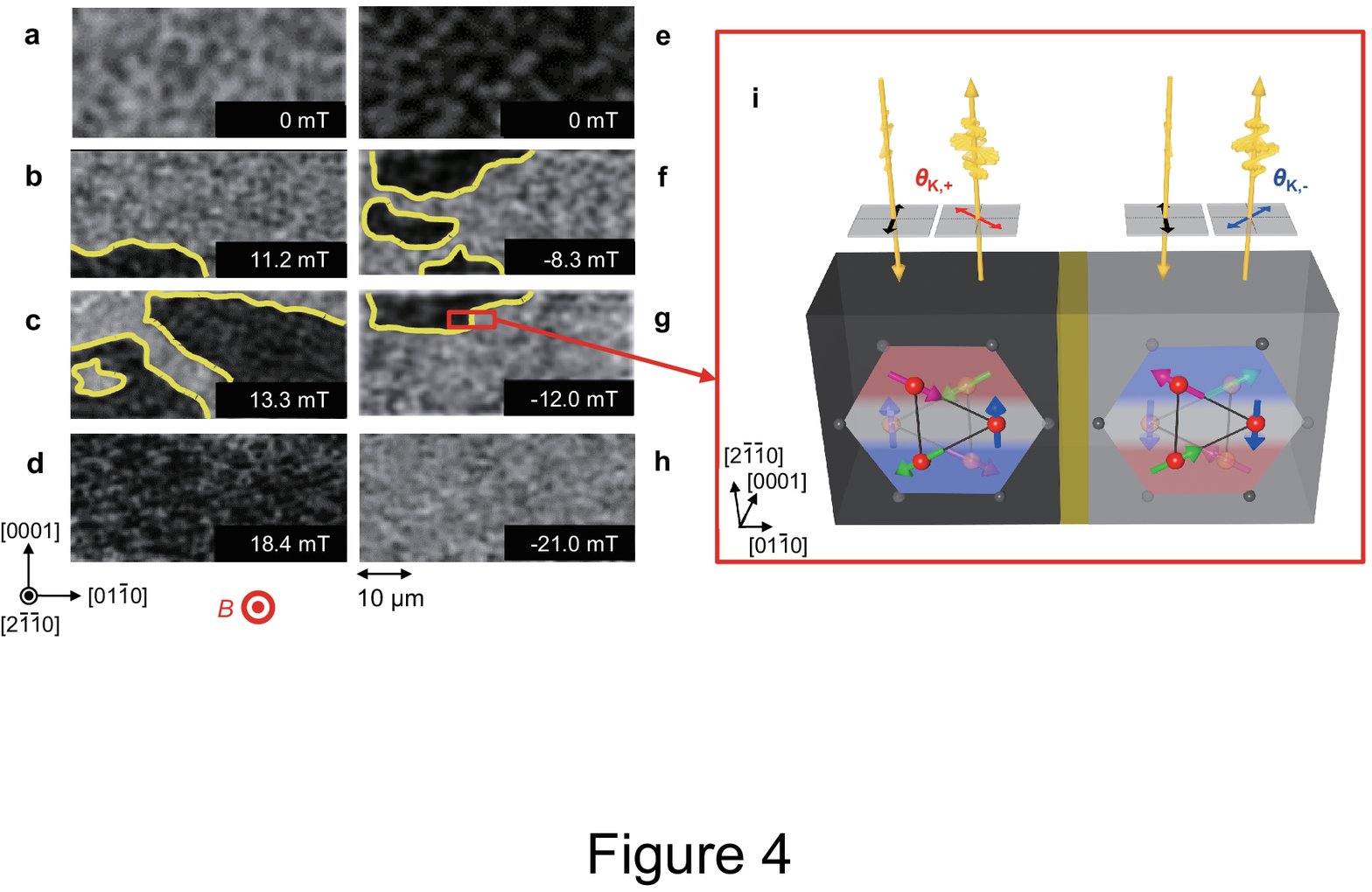}
\end{center}
\end{figure}

\end{document}